==================
\documentstyle[12pt]{article}

\let\de=\delta
\let\De=\Delta
\let\vareps=\varepsilon
\let\eps=\epsilon

\let\ka=\kappa

\let\p=\partial
\let\<=\langle
\let\>=\rangle

\def\comment#1{ \hbox{Comment suppressed here.} }

\makeatletter
\def\eqnarray{\stepcounter{equation}%
              \let\@currentlabel=\theequation
              \global\@eqnswtrue
              \global\@eqcnt\z@
              \tabskip\@centering
              \let\\=\@eqncr
              $$%
 \halign to \displaywidth\bgroup
    \eqnumphantom\@eqnsel\hskip\@centering
    $\displaystyle \tabskip\z@ {##}$%
    &\global\@eqcnt\@ne \hskip 2\arraycolsep
         \hfil$\displaystyle{##}$\hfil
    &\global\@eqcnt\tw@ \hskip 2\arraycolsep
         $\displaystyle\tabskip\z@{##}$\hfil
         \tabskip\@centering
    &{##}\tabskip\z@\cr}
\def\eqnumphantom{\phantom{(\theequation)}}

\newcommand{\ra}{\rightarrow}
\newcommand{\be}{\begin{equation}}
\newcommand{\ee}{\end{equation}}
\newcommand{\ba}{\begin{eqnarray}}
\newcommand{\non}{\nonumber \\}
\newcommand{\ea}{\end{eqnarray}}
\newcommand{\baa}{\begin{eqnarray*}}
\newcommand{\eaa}{\end{eqnarray*}}
\newcommand{\barr}{\begin{array}}
\newcommand{\earr}{\end{array}}
\newcommand{\bb}{\begin{thebibliography}}
\newcommand{\eb}{\end{thebibliography}}
\newcommand{\ci}[1]{\cite{#1}}
\newcommand{\bi}[1]{\bibitem{#1}}
\newcommand{\lab}[1]{\label{#1}}
\newcommand{\re}[1]{(\ref{#1})}

\newcommand{\PRL}[1]{~Phys.Rev.Lett.~{\bf #1}}
\newcommand{\PR}[1]{~Phys.Rev.~{\bf #1}}
\newcommand{\PL}[1]{~Phys.Lett~{\bf #1}}

\def\ltap{\raisebox{-.55ex}{\rlap{$\sim$}} \raisebox{.4ex}{$<$}}
\def\gtap{\raisebox{-.55ex}{\rlap{$\sim$}} \raisebox{.4ex}{$>$}}
\def\gsim{\mathrel{\gtap}}
\def\lsim{\mathrel{\ltap}}

\newcommand{\pg}{pairing~}
\newcommand{\eq}{$(++)$~}
\newcommand{\op}{$(+-)$~}
\newcommand{\AB}{Aharonov-Bohm~}

\begin{document}

\thispagestyle{empty}
\setcounter{page}{0}

\begin{flushright}
{\large UBCTP 92-7\\
March 1992}
\end{flushright}
\vspace{1cm}

\begin{center}
{\Large \bf Investigations of Pairing in Anyon Systems}\footnote{This work is
supported in part by the Natural Sciences and Engineering Research Council of
Canada} \\

\vspace{1.0cm}
{\large Mikhail I. Dobroliubov}\footnote{On leave from Institute for Nuclear
 Research, Moscow},
{\large Ian I. Kogan}\footnote{ On leave from ITEP, Moscow.},
  and {\large Gordon W. Semenoff}\\
\vspace{0.6 cm}
{\large \it Department of Physics, University of British Columbia\\
Vancouver, B.C., Canada V6T2A6}\\
\vspace{0.2in}
\vspace{0.3in}

{\bf Abstract} \\
\end{center}
\vspace*{-5mm}
\noindent

We investigate pairing instabilities in the Fermi-liquid-like state of
a single species of anyons.
We describe the anyons as Fermions interacting with a Chern-Simons gauge
field and consider the weak coupling limit where their statistics
approaches that of Fermions.  We show that, within the conventional BCS
approach, due to induced repulsive Coulomb and current-current interactions,
the attractive Aharonov-Bohm interaction is not sufficient to generate a gap
in the Fermion spectrum.

\newpage

The possibility of superconductivity in anyon systems is of general interest.
The mean field theory approach to systems with a single species of anyons
developed by Fetter, Hanna and Laughlin
and other authors \cite{anyon} argued that a finite anyon density
induces a background statistical magnetic field and a Landau level
picture of the ground state and the ensuing energy gap and superconductivity
emerged. There are also alternative scenarios for anyon superconductivity
where the background statistical magnetic field is absent, the ground
state of the anyon system resembles that of a Fermi
liquid and the superconductivity is realized in the conventional BCS mode.
This occurs in a system which contains {\it two species} of anyons of
opposite charges with respect to the statistical gauge field and which
therefore have opposite fractional statistics
\cite{kog1,kogkhvesh,Khleb,BalKalm,misha} (we denote this as $(+-)$ pairing).
There, the statistical background magnetic fields cancel and it has been
found that the Aharonov-Bohm interaction is attractive enough to generate
the pair formation.

Recently, it has been noted that systems with a {\it single species} of
anyons and a Fermi-liquid-like ground state are also of
interest\cite{wil}. Anyons are described as Fermions interacting
with a Chern-Simons gauge field so that in the weak coupling limit
their statistics approaches that of Fermions. The Fermi-liquid state
is achieved when the mean background statistical magnetic field is cancelled
by an {\it external} magnetic field. A pairing instability in such systems
(which we denote $(++)$ ) has been suggested as an explanation of
the even denominator states in the fractional quantum Hall effect.
It has been conjectured in \cite{wil} that the Aharonov-Bohm interactions
(with coupling constant $1/\ka$) which give the Fermions
fractional statistics are sufficient to drive a
superconducting pairing instability.
In this Letter we shall show that this is not the case.  When the
back-reaction of the gap in the Fermion spectrum to the interaction
potential is taken into account, the BCS gap equation has only the
trivial solution.  We conclude that, if there is such an instability
in this system, it must occur outside of the weak-coupling BCS scenario
or else be driven by other interactions. It is also worth noting that in
the physical systems of interest
$\ka$ is small (for example in \cite{wil} $\ka=1/2\pi$), and {\it strong
coupling} effects could lead to the gap formation (in that scenario there
would be a critical $\ka$  where pairing takes place).  In that case,
we observe that (as was previously found for the \op case \ci{misha}) the
coefficient of the Chern-Simons term is renormalized as $\ka\rightarrow
\ka-\ell/4\pi$,  where $\ell$ is the angular momentum of the Cooper pair
which, because of Fermi statistics, must be an odd integer.
Thus, the Hall conductivity and fractional statistics of quasiparticles
and magnetic vortices are modified by the gap.  Due to no-renormalization
beyond one loop arguments \cite{col} for the Chern-Simons term we expect
that the
latter result is valid beyond our large $\ka$ perturbation theory.
To get an even denominator Hall conductance, $2\pi k$ should be a
half-odd integer.

It is interesting to compare the \eq case with the \op case where it has
been shown \cite{misha} that Aharonov-Bohm interactions do lead to pairing
and formation of an energy gap, which even proves to be parametrically larger
than the simplest BCS gap. At the tree level these cases are equivalent:
both contain only long-ranged \AB interaction (which might be both attractive
and repulsive, depending on the angular momentum of the anyon pair
\ci{Khleb,misha}). Bare
Coulomb and magnetic current-current interactions are absent, but will be
generated by the radiative corrections. The difference between the \op and
\eq cases stems from different back-reactions of the gap formation on the
effective interaction between anyons.

	Since the \op pairs are neutral with respect to the statistical
gauge field, their presence in the ground state does not lead to
 a Meissner effect (i.e. generation of a London mass) for the statistical
gauge field. Moreover, since at zero temperature all anyons are bound in
pairs, there are no free charge carriers in the
system, and therefore, no Debye screening. This in turn results in the
fact that radiatively induced Coulomb and magnetic current-current interactions
are short ranged (topologically massive), whereas the Aharonov-Bohm
interaction remains long ranged and, therefore, dominant
 \ci{misha}.  In the equal charge case, on the contrary, the formation
 of \eq pairs of course leads to both Debye and Meissner screening,
and {\it all} interactions become short ranged. Thus the interaction
 which causes anyon pairing will in turn be completely changed by
 this very pairing. Note that the values of the Debye and London masses are
defined only by the density of the free charge carrers in the former case
and that of the superconducting fermions in the latter one. If there is a gap
in the Fermionic spectrum these two are defined by the density of anyons
in the system, and for zero temperature have nothing to do with the magnitude
of this gap, irrespectively how small it is.  They can therefore have an
important influence on the gap equation. The purpose of this paper is to
consider a self-consistent picture of this feedback. We will demonstrate
that for large $k$ ($2\pi/k$ is the statistics parameter) there is no gap in
the Fermionic spectrum, contrary to the  result in \cite{wil}. For simplicity
we will consider the zero-temperature case.

We start with the lagrangian
\begin{equation}
L=-\frac{\kappa}{2}\epsilon^{\mu\nu\lambda} A_{\mu}\partial_{\nu} A_{\lambda} +
\Psi^{\dagger}\left( i\partial_0 - A_0 -
\epsilon\left( -i\frac{\partial}{\partial {\bf x}} - {\bf A}
\right)\right)\Psi~~~.
\label{lag}
\end{equation}
where the statistical coupling $\kappa$ is taken positive and large,
and $\epsilon({\bf k})$ is the (quasi)par\-tic\-le (anyon) dispersion law below
taken to be $\epsilon({\bf k})={\bf k}^2/2m -\epsilon_F$. The tree-level
potential has pure \AB form and reads in the radiation gauge ($\p_i A_i =0$)
\begin{equation}
U({\bf p,p'}) = {2i\over\kappa m}~
{\varepsilon_{ij} p_ip'_j\over ({\bf p-p'})^2} = -
  {2i\over\kappa m}~
  \frac{\mbox{sin}\theta}{{p\over p'} + {p'\over p} -
2\mbox{cos}\theta} \; ,
\label{U}
\end{equation}
where ${\bf p}=p (\cos\phi, \sin\phi)$,
{}~${\bf p'}=p' (\cos\phi', \sin\phi')$,
{}~$\theta=\phi-\phi'$.
This potential leads to the formation of bound
pairs of anyons with non-zero angular momentum \ci{Khleb}. Substituted
into the standard  BCS gap equation
\be
\Delta_{\bf p}=-\frac{1}{2}\int \frac{d{\bf p'}}{(2\pi)^2}~ U_{{\bf
p}{\bf p'}} \frac{\Delta_{\bf p'}}{\sqrt{\epsilon_{\bf p'}^2 +
|\Delta_{\bf p'}|^2}} ~~,
\label{gap}
\ee
it gives rise to the exponentially suppressed gap, which depends nontrivially
on the momentum \ci{misha}
\be
\Delta ({\bf p}) = \Delta e^{i\ell \phi} \left[
\left( \frac{p}{p_F} \right)^\ell \theta (p_F-p) +
\left( \frac{p_F}{p} \right)^\ell \theta (p-p_F) \right] ~~.
\lab{ansatz}
\ee
The largest gap is achieved for P-wave pairing
\be
\Delta \simeq \eps_F e^{-2\pi\kappa}~~~.
\lab{bare gap}
\ee

Taking into account the back-reaction of the gap on the
renormalization of the potential leads to drastic changes in these
tree-level conclusions. In the case of \op \pg it removes the
exponential suppression of the gap \ci{misha}
\begin{equation}
\Delta\sim\frac{\epsilon_F}{\kappa} ~,
\label{AB}
\end{equation}
which demonstrates the existence of a new universality class for
the superconducting order parameter.

Now we will show that in the \eq case the result is quite the opposite:
the only solution of the renormalized gap equation is $\Delta =0$.

To find the gap improved potential we need to know the polarization
operator (herewith we will be interested in the static limit, when
there is no energy transfer through the gauge line, and we will
restrict ourselves to the first loop correction, as in the large
$\kappa$ limit the contribution of higher loops will be suppressed)
. In the radiation gauge we find \ci{to be published}

\noindent
{\bf 1)} $\Pi_{00}({\bf q})= i \frac{m}{2\pi}$, for all  $|{\bf q}|<2p_F$.
Indeed, in the case of the single charge plasma the Debye screening
does not depend on whether or not the Fermions are bound in the pairs.

\noindent
{\bf 2)} $\Pi_{ij}({\bf q})= -i \Pi({\bf q}) \delta_{ij}~.~\Pi({\bf q}=0)=
\frac{\eps_F}{4\pi} \delta_{ij}$. The nonzero value of
$\Pi_{ij}({\bf q}=0)$ indicates the appearance of the magnetic London mass of
the statistical photon (Meissner effect). Note that this London mass
does not depend on the value of the gap, exhibiting thus a nonanalytic
behavior, as for the case of the absent gap $\Pi ({\bf q}=0)=0$.
This is of no surprise, since we know that the value of the London
mass is determined by the density of superconducting Fermions, and at
 zero temperature all Fermions are superconducting however small
the gap.  When $q=|{\bf q}|$ increases $\Pi (q)$ decreases, and
 for large $q \gg \Delta/v_F$ it approximately equals $\Pi (q)
\simeq (\frac{\Delta \eps_F}{q v_F} + \frac{q^2}{24\pi m})
\delta_{ij}$, where the last term in the brackets is the magnetic
response of the normal metal.

\noindent
{\bf 3)} $\Pi_{0i}({\bf q}=0)=  \frac{\ell }{4\pi} \vareps_{ij}q_j$,
where $\ell $ is the
angular momentum of the pair. This leads to the same renormalization
of the Chern-Simons coupling as those first found in \ci{misha}
$\kappa_{ren}=\kappa-\ell /4\pi$, but it is worth recalling that now,
after the formation of the BCS pairs, the \AB interaction is short
ranged, and this renormalization does not lead to a considerable enhancement of
the \AB attraction and in what follows we will neglect it in the
large $\kappa$ limit. Since there is no corrections to the
Chern-Simons term in the ordinary free plasma, for $q \gg \Delta/v_F$
$\Pi_{0i}$ decreases to zero.

The full gauge propagator is in the matrix notations
\be
D^{-1} = D_0^{-1} - P~~~,
\ee
where $D_0$ is the bare propagator, the only nonzero component of
which is
\be
D_{0i} ={1\over\kappa} \frac{\vareps_{ij} q_j}{q^2} ~~,
\ee
and
$P$ is the polarization operator. So,
\ba
D_{00} = i D_0 ~~,~~ D_0 = \frac{\Pi}{\kappa^2 q^2 + \frac{m\Pi}{2\pi} }
{}~~,
\non
D_{ij} = - i D_1 \left( \de_{ij} - \frac{q_i q_j}{q^2}
\right) ~~,~~ D_1 = \frac{m/2\pi}{\kappa^2 q^2 + \frac{m\Pi}{2\pi} }
{}~~,
\non
D_{0i} = D_2 \vareps_{ij} q_j ~~,~~ D_2 = \frac{1}{\kappa
q^2 + \frac{m\Pi}{2\pi\kappa}} ~~.
\ea

The behavior of the functions $D_{0,1,2}$ is depicted on Fig.1. Note, that
for
\be
q\lsim q_0 \simeq \left( \frac{\De}{\ka^2\eps_F} \right)^{1/3}~~,
p_F
\lab{q0}
\ee
the propagators $D_{1,2}$ {\it increase} with increasing $q$. This very
unusual behavior of a propagator would be forbidden in the
relativistic case, and is a net effect of the finite density of
anyons, presence of the gap in their spectrum, and the form of the
bare gauge propagator (domination of the Chern-Simons term in it).  It
makes the most important region of the transferred momenta $q^2=({\bf p} -
{\bf p'})^2$ in the gap equation \re{gap} to be $q\simeq q_0 $, as
opposed to the standard case where this region was $q\simeq ({\De\over
\eps_F}) p_F \ll q_0$.

Note, that despite all our interactions are short ranged in the sense,
that all $D_i$ are finite (and non-zero) at $q=0$, the magnetic
interaction still contains the factor of  $1/q^2$ from the tensor
structure $D_{ij} = i D_1 \left( \de_{ij} - \frac{q_i q_j}{q^2}
\right) $ (this is a reminiscent of the gauge symmetry of the original
lagrangian \re{lag}). Therefore, the corresponding potential for the
current-current interaction in the $\ell $-wave pairing (cf.
(\ref{U},\ref{ansatz}) )
\be
U_M ({\bf p,p'}) = {i\over m^2} p_i p_j D_{ij} \cos {\ell \theta} =
{1\over m^2} D_1 \frac{p^2 p'^2 \sin {\theta^2} \cos {\ell \theta}}
{p^2+p'^2-2pp'\cos {\theta}} ~~,
\ee
is finite when ${\bf p'}\ra{\bf p}$ ($p'\ra p,~\theta\ra 0$).

The two others interactions, Coulomb
\[
U_C = -i D_{00} \cos{\ell \theta} = D_{0} \cos{\ell \theta}
\]
and \AB,
\[
U_{AB} = -\frac{2ip_i}{m} D_{0i} \, i \sin{\ell \theta} =
-{2\over m} D_2 \frac{p p' \sin{\theta}\sin{\ell \theta}}
{p^2+p'^2-2pp'\cos {\theta}} ~~,
\]
 vanish in this limit, and are,
therefore, subdominant with respect to the current-current
interaction.

Let us stress that taking into account the back-reaction of \pg onto
the potential between anyons results in weaking of the attractive part of
the potential, and appearing of the strong repulsive part. So we conclude,
that the gap should be suppressed even stronger that it was at the tree
level (see \re{bare gap} ) and the ratio $\De/\eps_F$ is the smallest parameter
 in the theory.

It can be seen that parametrically the RPA improved gap equation reads
\ci{to be published}
\be
\De = \left\{ -\frac{\De^{2/3} \eps_F^{1/3}}{\ka^{4/3}} -
\frac{\De^{4/3}}{\ka^{4/3} \eps_F^{1/3}}
+ {\De\over \ka} \right\} \log \left( {\eps_F\over\De} \right) ~~,
\lab{new gap}
\ee
where the first term in the curly brackets is the contribution of the
magnetic current-current interaction, the second one is that of the Coulomb
interaction (these two come from the small  momentum transfer in the gap
equation \re{gap}, $|{\bf p} - {\bf p'}| \sim q_0$)
, and the last one is the contribution of the \AB interaction
(it comes from the large momentum transfer in the gap equation,
$|{\bf p} - {\bf p'}| \gg q_0$ for which it has essentially the bare form).
We see, that only the trivial solution $\De =0$ satisfies \re{new gap},
provided the constraint $\De\lsim \eps_F e^{-2\pi \kappa}$ is fulfilled.
Therefore we conclude,
that within the simplest BCS picture there is no pairing in the Fermi
liquid of anyons of the same charge.

It also seems unlikely that for \pg at large angular momenta $\ell$ the system
might become  superconducting due to the Kohn-Luttinger effect \ci{KL}.
Indeed, as we showed, the most important region of the transferred
momentum in the gap equation \re{gap} is $q\sim q_0$, where the net
interaction is repulsive, what makes the improved gap equation \re{new gap} to
have no non-trivial solutions. To smear out this repulsion one
needs very large angular momenta,
\be
\ell \gsim \left( \frac{\ka^2\eps_F}{\De} \right)^{1/3}~~,
\ee
which look implausible.

The arguments presented above, of course, do not rule out the general
possibility of the superconductivity in these systems. There might be
other, more complicated mechanisms of \pg, like gapless
superconductivity \ci{cohen}. Besides that, as we mentioned above,
for real physical
systems the value of the statistics parameter $\ka$ is rather small, so
these systems correspond to the strong coupling regime.
For example, for $\ka \sim 2\pi$ the statistics of
anyons is close to that of bosons, and the system of equally charged anyons
in this case will behave like the XY model, for which the long-range order
exists \ci{XY}. All these should be a subject of further investigations.

The authors are grateful to D.Eliezer, S.Yu.Khlebnikov, P.C.E.Stamp and
N.Weiss for numerous valuable discussions.

\newpage

\begin{center}
{\bf FIGURE CAPTION}
\end{center}

\noindent {\bf a)}  The behavior of the propagator $D_{00} (q)=iD_0(q)$:\\
\noindent {\it 1)} $q \ll q_0$, ~$D_0 \simeq \frac{2\pi}{m}$\\
\noindent {\it 2)} $q_0\ll q \ll q_2$, ~$D_0 \simeq \frac{1}{4\ka^2 m}
\frac{\De}{\eps_F} \frac{p_F^3}{q^3}$\\
\noindent {\it 3)}  $q_2 \ll q \lsim p_F$, ~$D_0 \simeq
\frac{1}{24\pi\ka^2 m}$ \\

\noindent {\bf b)} The behavior of the propagators
$D_{ij}(q)=-iD_1(q) \left( \delta_{ij} -\frac{q_iq_j}{q^2} \right)$,
{}~$D_{0i}(q)=D_2(q) \vareps_{ij} q_j$:
\noindent {\it 1)}  $q\ll q_1$, ~$D_1 \simeq \frac{4\pi}{\eps_F}$,
{}~$D_2\simeq \frac{16\pi^2}{p_F^2}$\\
\noindent {\it 2)}  $q_1\ll q \ll q_0$, ~$D_1 \simeq \frac{1}{\eps_F}
\frac{2 \eps_F}{\De} \frac{q}{p_F}$, ~$D_2 \simeq \frac{1}{p_F^2}
\frac{16\pi\ka\eps_F}{\De} \frac{q}{p_F}$\\
\noindent {\it 3)}  $q_2 \ll q \lsim p_F$, ~$D_1 \simeq
\frac{1}{4\pi\ka^2\eps_F} \frac{p_F^2}{q^2}$, ~$D_2 \simeq
\frac{1}{\ka p_F^2} \frac{p_F^2}{q^2}$\\
\noindent Maximal values $D_1 (q_0) \sim \frac{1}{\eps_F}
\left( \frac{\eps_F}{\ka\De} \right)^{1/3}$, ~$D_2 (q_0) \sim \frac{1}{p_F^2}
\left( \frac{\ka^2\eps_F}{\De} \right)^{1/3}$ \\

\noindent The values of the crossover momenta are: $q_0 \sim
\left( \frac{\eps_F}{\ka^2\De} \right)^{1/3} p_F$,
$~q_1 \sim \left( \frac{\eps_F}{\De} \right) p_F$,\\
$q_2 \sim \left( \frac{\eps_F}{\De} \right)^{1/3} p_F$.

\newpage
\begin{thebibliography}{99}

\bibitem{anyon} A.Fetter, C.Hanna and R.B.Laughlin, Phys. Rev.
{\bf B39}, 9679 (1989); Y.H.Chen, F.Wilczek, E.Witten and
B.I.Halperin, Int.J. Mod. Phys. {\bf B3}, 1001 (1989);

\bibitem{kog1} I.I.Kogan, JETP. Lett. {\bf 49}, 225 (1989)

\bibitem{kogkhvesh} D.V.Khveshchenko and I.I.Kogan, Mod. Phys. Lett.
 {\bf B4}, 95 (1990), Int. J. Mod. Phys. {\bf B4}, 631, 1990

\bibitem{Khleb} S.Yu.Khlebnikov, \PR{B43}, 2381 (1991)

\bibitem{BalKalm} A.Balatsky and V.Kalmeyer, Phys. Rev., {\bf
B43}, 6228 (1991)

\bi{misha} M.Dobroliubov, S.Khlebnikov, \PRL{67}, 2084 (1991)

\bi{wil} M.Greiter, X.Wen, F.Wilczek, \PRL{66}, 3205 (1991);
{\it "On Paired Hall States"}, IAS preprint IASSNS-HEP-91/66;
{\it "Paired Hall States in Double Layer Electron Systems"},
IAS preprint IASSNS-HEP-92/1

\bi{col} S.Coleman and B.Hill, \PL{159B}, 184 (1985);
G.Semenoff, P.Sodano and Y.-S.Wu, \PRL{62}, 715 (1989)

\bi{to be published} the details will be presented elsewhere

\bi{KL} W.Kohn, J.M.Luttinger, \PRL{15}, 524 (1965)

\bi{cohen} M.H.Cohen, \PRL{12},664 (1964)

\bi{XY} T.Kennedy, E.H.Lieb, B.S.Shastry, \PRL{61}, 2582 (1988)

\end{thebibliography}
\end{document}